\documentclass[]{iau}
 \usepackage{graphicx,subfigure} 

\title[IAUS291.~~Pulsars in gamma rays] 
{Pulsars in gamma rays: \\ What {\slshape Fermi} is teaching us} 

\author[M. Kerr]  
{Matthew Kerr$^1$
 \and the Fermi-LAT Collaboration}

\affiliation{$^1$Kavli Institute for Particle Astrophysics and Cosmology, Stanford University, \\Stanford, CA 94305, USA \\ email: {\tt kerrm@stanford.edu} \\[\affilskip]
}

\pubyear{2012}
\volume{291}  
\jname{\mbox{Neutron Stars and Pulsars: Challenges and Opportunities after 80 years}}
\editors{J. van Leeuwen, ed.} 
\begin{document}

\maketitle

\begin{abstract}
The 2nd {\it Fermi}-LAT pulsar catalog includes 117
$\gamma$-ray pulsars, of which roughly one third are millisecond
pulsars (MSPs) while the remaining two thirds split evenly into
young radio-loud and radio-quiet pulsars.  Although this large
population will enable future, detailed studies of emission mechanisms
and the evolution of the underlying neutron star population, some
nearly-universal properties are already clear and
unequivocal.  We discuss some of these aspects below, including the
altitude of the $\gamma$-ray emission site, the shape of the
$\gamma$-ray spectrum, and the  implications of the latter for the radiation
mechanism. 
\end{abstract}


\firstsection 
\section{Introduction}

Since their discovery forty-five years ago (\cite[Hewish et al.
1968]{hewish68}), pulsars have primarily been studied at radio
wavelengths---a natural consequence of their discovery mode and the
progression to ever larger antennae and higher-bandwidth receivers.
Yet pulsars emit only a tiny fraction ($\sim$$10^{-5}$) of their
spindown luminosity $\dot{E}$ at radio wavelengths.  Furthermore,
their high brightness temperature and variability on timescales as
short as a few ns (Hankins et al. 2003\cite[]{hankins+_2003}) indicate
a coherent emission process from a small volume.  Making inferences
about the global structure of the pulsar magnetosphere from such local
measurements is difficult.

\begin{figure}[b]
 \vspace{0.5 cm}
\begin{center}
\hspace{-3.0 cm}
\includegraphics[width=3.4in,scale=0.4]{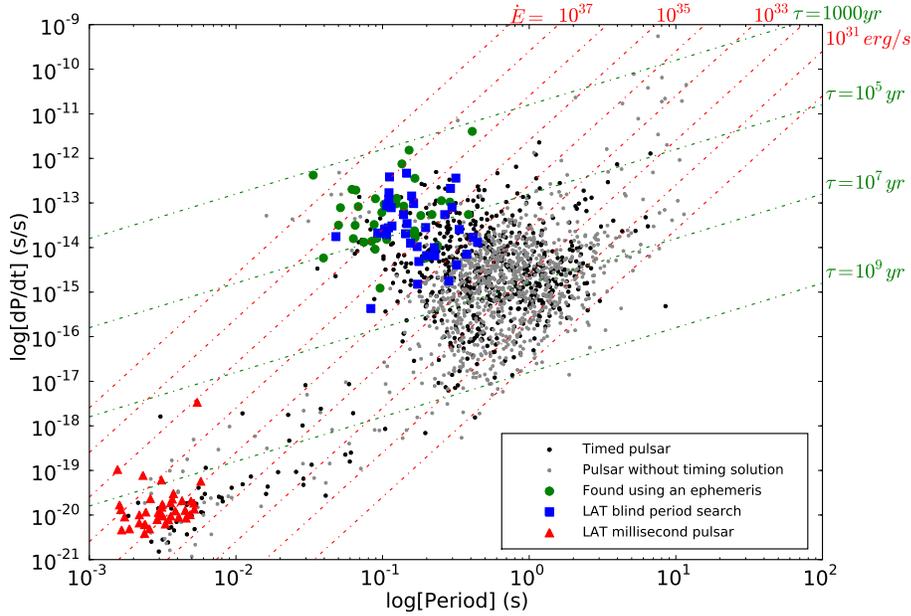}%
 \caption{The period (P) and period derivative (dP/dt) of 2PC pulsars.}
   \label{fig_ppdot}
\end{center}
\end{figure}

Young pulsars and MSPs, however, are extraordinarily efficient sources
of $\gamma$ rays, converting  1--100\% of their spindown
luminosity into 100\,MeV--30\,GeV emission (e.g. Abdo et al.
2010b, and noting that efficiencies can be highly
uncertain due to anisotropic beams and distance uncertainty).  The
emission is incoherent, and such efficiency implies much of the
potential induced by the pulsar's magnetic field is involved in
particle acceleration, i.e. the acceleration and emitting volumes are
large on the scale of the magnetosphere.  Moreover, the
ultrarelativistic particles powering the $\gamma$ emission propagate
along magnetic field lines and beam their emission along them.  Thus,
$\gamma$ rays are a tracer of the large scale magnetic field
structure.

Prior to the {\it Fermi} Large Area Telescope (LAT), only a handful of
$\gamma$-ray pulsars were known, primarily of the young, radio-loud
class, limiting their usefulness in studies of the magnetosphere and
the underlying pulsar population.  The situation evolved rapidly after
the successful launch of {\it Fermi}, with the discovery of
$\gamma$-ray emission from MSPs (Abdo et al. 2009a) and radio-quiet
young pulsars from fecund ``blind'' searches of the LAT data (Abdo et
al. 2009b).  The first {\it Fermi} pulsar catalog (Abdo et al. 2010b)
provided properties of 46 $\gamma$-ray pulsars, an appreciable
increase in number and diversity.  The second {\it Fermi} pulsar
catalog (2PC), in preparation, includes 117 $\gamma$-ray pulsars and
more sophisticated analysis.  For instance, light curves use photon
weights (Kerr 2011) which provide increased S/N and a
reliably-estimated background level.  In-depth spectral analyses
attempt to disentangle magnetospheric emission from astrophysical
backgrounds, including pulsar wind nebulae.

2PC will undoubtedly fuel detailed population (synthesis) studies.
The breadth and depth of the observed population, however, suggest
that any observed trends are nigh-universal properties of $\gamma$-ray
pulsars and are worth studying in their own light.  Below, we detail
two such trends.  First, $\gamma$-ray emission appears to originate in
the outer magnetosphere.  Second, the spectral properties of the
emission, for all pulsars save the Crab, are consistent with curvature
radiation.

\section{$\gamma$-ray emission altitude}

The height in the magnetosphere from which the bulk of
$\gamma$-ray emission arises is probed through several mechanisms.
The first, relatively model-independent method, depends on the
altitude-dependent opacity for pair production by $\gamma$-rays on the
strong magnetic field near the surface of the neutron star.  Two
additional methods depend on the assumed structure of the magnetic
field, but with little sensitivity to the details.  Below, we briefly
discuss each method.

\subsection{Magnetic opacity}

In strong magnetic fields, classically forbidden processes, such as
one-photon pair production and annihilation, become allowed (e.g.
Harding 1991).  The former process, in particular, attenuates
high-energy $\gamma$ rays.  The altitude by which the magnetic field
has weakened sufficiently to allow photons of energy $E$ to escape is
given by $r\approx(B_{12}\,E/1.76\,{\rm GeV})^{2/7}P^{−1/7}R_{\star}$
(Baring 2004) with $B_{12}$ the magnetic field in units of $10^{12}$G
and $R_{\star}$ the neutron star radius.  The observation of photons
well above 10\,GeV for young neutron stars implies production above a
few $R_{\star}$.  Although this constraint is small in terms of the
distance to the light cylinder, $R_{LC}$ (which depends on period, but
for young pulsars is $\sim$200--2000$\,R_{\star}$), it rules out
emission from particles accelerated near the polar cap (e.g. Arons and
Scharlemann 1979).

Additional evidence along these lines stems from the shape of the
spectral cutoff observed in {\it Fermi} pulsars.  Since the
opacity to magnetic attenuation depends on energy, the curvature
radiation spectrum from low-altitude photons is suppressed
superexponentially.  The observed spectra, however, show no such
suppression.  Indeed, as discussed below, they typically decay
subexponentially; see, e.g. the spectra of Geminga (Abdo et al. 2010a)
and Vela (Figure \ref{fig_vela_sed}).

\subsection{Caustics}

Although observations of high energy ($\gg1\,$GeV) photons and gradual
spectral cutoffs rule out emission from near the stellar surface, they
do not help to distinguish between ``low altitude'' (10--30\% of
$R_{LC}$) and ``outer magnetospheric'' (OM; $>$30\%$R_{LC}$) origin.
These definitions are somewhat arbitrary, and another convenient
division between low altitude and OM emission is the null charge (NC)
surface, where the magnetic field projected along the spin axis
vanishes.  Indeed, the NC surface is the traditional starward edge of
the eponymous ``outer gap'' of such particle acceleration models (e.g.
Cheng, Ho, and Ruderman 1986).  Other models, such as the slot gap
(e.g. Muslimov and Harding 2003), predict emission both above and
below the NC surface.

A general property of emission from the OM is the formation of
caustics from near cancellation of relativistic aberration, time of
flight across the magnetosphere, and field line sweepback (Morini
1983).  The degree of sweepback determines the precise location in the
magnetosphere for caustic formation, but static, vacuum, and
force-free MHD magnetic field solutions all demonstrate such features,
as shown e.g.  by Romani and Yadigaroglu (1995), Watters et al.
(2009), and Bai and Spitkovsky (2010), respectively.  In the
population of young 2PC pulsars, caustic-like peaks are a nearly
universal light curve feature, a clear indication of OM origin.

For radio-loud pulsars, two important quantities may be estimated:
$\delta$, the phase lag from the magnetic axis to the first $\gamma$
peak, and $\Delta$, the separation of (if present) the two primary
$\gamma$ peaks.  Generally, OM magnetosphere and emission models
predict a correlation of these two quantities, with the shape
depending on the model details.  The observed correlation of 1PC
agrees roughly with ``outer gap'' models (see e.g., Watters et
al. 2009) and the correlation is strengthened for young pulsars in
2PC. 
The precise phase relation is a sensitive probe of the structure of
the magnetosphere and may reveal, e.g., the extent to which currents
shape the magnetosphere.  MSPs show a similar, though more scattered,
correlation, with some exceptions.  These are ``aligned'' MSPs (e.g.
PSR B1937+21, see Guillemot et al. 2012) whose $\gamma$-ray and radio
emission occur at the same rotational phase, inviting speculation that
both beams originate from the same OM caustics.

\subsection{Unpulsed magnetospheric emission}

A final argument for the dominance of OM $\gamma$-ray emission is the
general absence of unpulsed emission---that is, emission present
throughout all rotational phases.  The level of such emission is
robustly estimated in 2PC through the use of photon weights; these
weights are derived directly from the phase-averaged spectral model,
which accounts for the spatial distribution of astrophysical
backgrounds.  In contrast, background levels in 1PC were estimated by
considering photon counts in an off-pulse annulus around the pulsar,
subject to substantial inaccuracy from spatial variation in the
background level.

In the altitude range above the polar cap but below the outer
magnetosphere, the divergence of field lines causes beams from the
open zone to illuminate an ever greater fraction of the sky with
increasing altitude, leading to emission with little
modulation at sufficiently high altitudes.  At even higher (OM)
altitudes, caustics begin to form, leading to anisotropic and
fully-pulsed emission.  Thus, the presence of unpulsed magnetospheric
emission is a sensitive indicator of emission from low altitudes.

In the 2PC, only two pulsars (PSRs J1836+5926 and J2021+4026) have
levels of unpulsed emission competitive with or exceeding modulated
emission, while the remaining pulsars show subdominant or absent
off-peak emission, suggesting that emission from low altitude plays a
minimal role in most $\gamma$-ray pulsars.

\section{$\gamma$-ray emission mechanism}

A second question 2PC is poised to answer is the nature of the
dominant $\gamma$-ray radiation mechanism.  The simplest, and arguably
most widely favored, process for producing the GeV photons is
curvature radiation from leptons with Lorentz factors of $\sim$10$^7$
flowing along magnetic field lines.  The primary alternative is
inverse Compton emission, potentially from synchrotron seed photons
(synchrotron/self-Compton).  A detailed review of these models is
beyond the scope of this paper.

The primary discriminator between the two mechanisms is the spectral
shape of the broadband emission.  While a few bright pulsars are
detected in hard X-rays and soft $\gamma$ rays, the LAT GeV spectrum
is the primary observable.  Although the expected spectral shape is a
complicated function of the acceleration and cooling of the emitting
particles, the spectral cutoff for curvature radiation is cleanly
determined by the highest-energy particles, and the absence of a
strong cutoff could be interpreted as evidence for IC emission.
However, as we show below, two effects---one technical and one
physical---confound such simple conclusions, and the vast majority of
{\it Fermi} pulsar emission is consistent with curvature radiation.

\subsection{Monoenergetic spectrum}

The curvature radiation spectrum of a single particle is set by the
particle energy and the radius of curvature of the magnetic field.
This monoenergetic photon spectrum, summed over polarizations, takes the simple
form (Jackson 1998)
\begin{equation}
\frac{dN}{dE} \propto \int_{E/E_c}^{\infty} K_{5/3}(x)\,dx,
\end{equation}
where $K$ is a modified Bessel function.  In the limits $E\ll E_c$ and
$E\gg E_c$, the spectrum asymptotes to the well known forms
$dN/dE\propto E^{-2/3}$ and $dN/dE\propto E^{-1/2}\exp -E/E_c$,
respectively.  Typical measured and expected values of $E_c$ of
1--3\,GeV fall squarely in the middle of the LAT passband.  The
functional form most often used to fit LAT spectra, $dN/dE\propto
E^{-\Gamma}\exp-E/E_c$ (PLEC), can only mimic one or the other of
these two limits.  Thus one must be cautious in interpreting a failure
of the PLEC model to fully describe the data as a failure for the
curvature radiation mechanism.

\begin{figure}[t]
\centering
\includegraphics[trim=20 0 40 30, clip, width=0.5\textwidth]{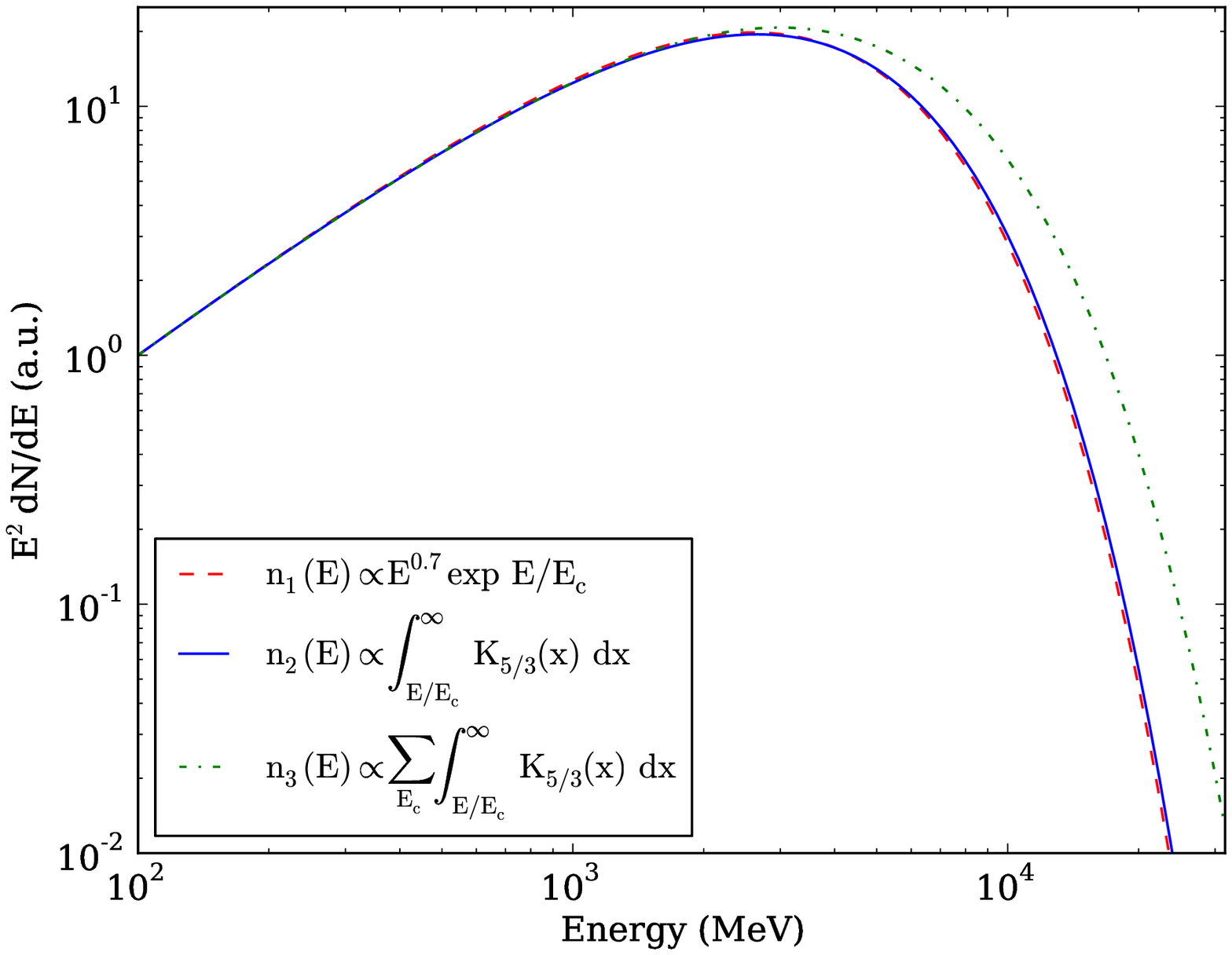}%
\includegraphics[trim=20 0 40 30, clip, width=0.5\textwidth]{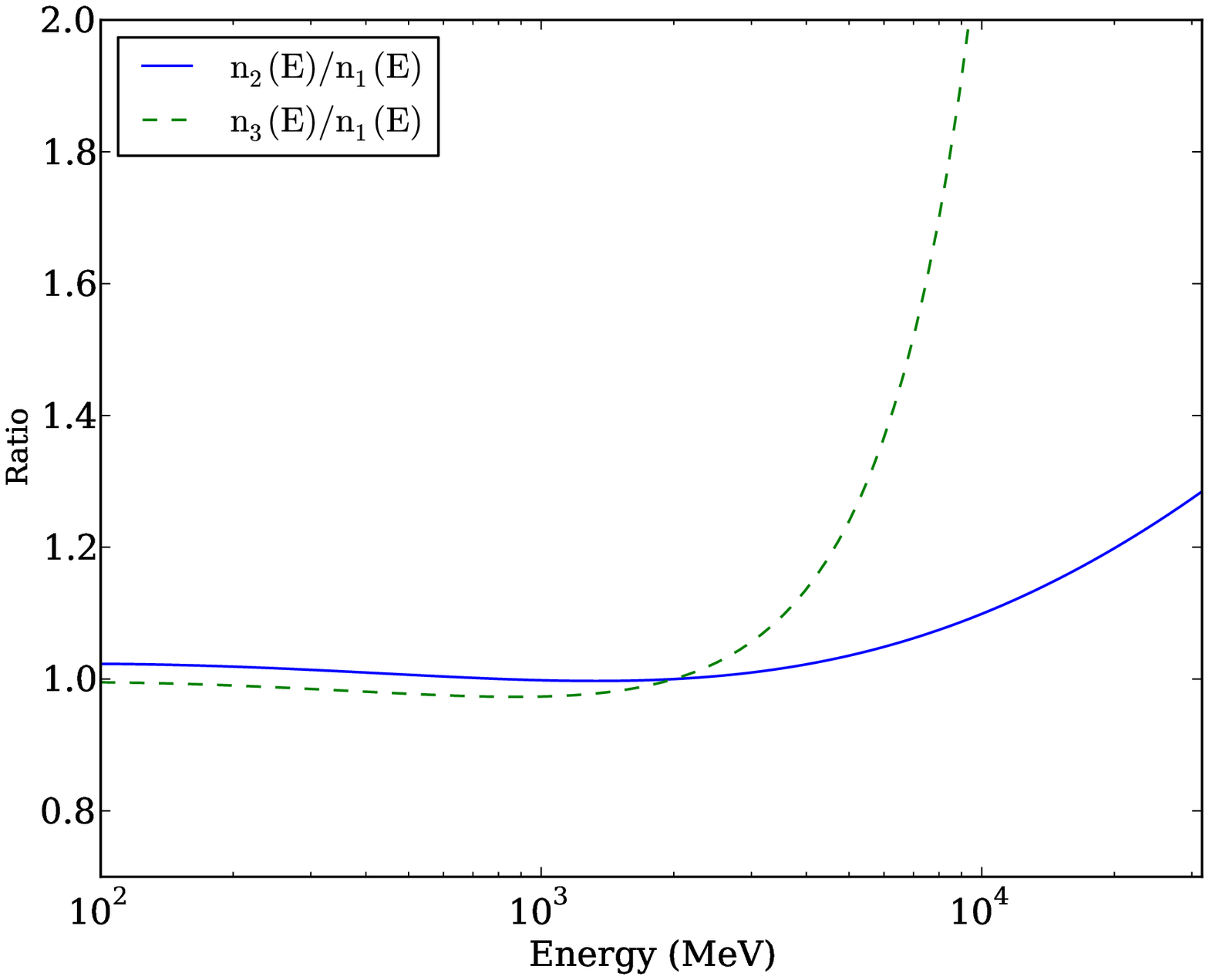}
\caption{{\it Left:} A monoenergetic ($E_c=2\,$GeV) curvature radiation
spectrum is shown in solid blue, along with a PLEC model that agrees
closely with it below $E_c$ (dashed red).  A mixture of monoenergetic
spectra appears in dot-dashed green. {\it Right:} The ratio of monoenergetic
spectra to a PLEC approximation.  A single spectrum ($E_c=2\,$GeV)
appears in solid blue, while the mixture spectrum described in the
text is shown in dashed green.}
\label{fig1}
\end{figure}

\begin{figure}[b]
\begin{center}
\includegraphics[trim=15 5 50 25, clip, width=4.1in]{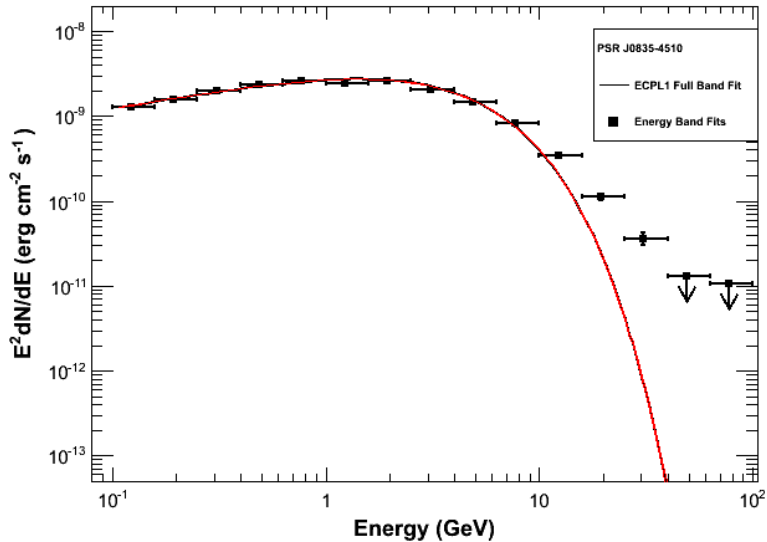}%
 \caption{The 2PC phase-averaged spectrum of PSR J0835-4510 (Vela).
 The best fit PLEC is shown as the solid, red line and substantially
 underestimates the observed spectrum.}
   \label{fig_vela_sed}
\end{center}
\end{figure}

We briefly probe the magnitude of the error made in approximating a
monoenergetic curvature spectrum with the PLEC model.  In Figure
\ref{fig1} (left), we show the monoenergetic photon spectrum with
$E_c=2\,$GeV.  The PLEC spectrum that most closely agrees at energies
$E<E_c$ with the exact curvature spectrum has $\Gamma=0.7$, close to
the expected asymptotic value 2/3.  However, this photon index is
softer than the high-energy value of 1/2, causing the PLEC model to
appreciably underestimate the monoenergetic spectrum for $E>E_c$,
resulting in a discrepancy of $\sim$20\% at 20 GeV (Figure \ref{fig1},
right).  Such energies are well within reach of the LAT, and indeed
the observed spectra of bright pulsars such as Vela show a
disagreement of just such a magnitude.  These conclusions are in line
with those of the phase-resolved spectroscopy of Abdo et al. (2010c)
where spectra in narrow phase windows agree well with a PLEC
model.

\subsection{Spectral superpositions}

An even more drastic effect stems from the caustic nature of OM
emission.  Because emission piles up over a large volume of the
magnetosphere, multiple field lines / radii of curvature are expected
to contribute to the phase-averaged spectrum.  For a constant
accelerating electric field, this leads to a linear spread in $E_c$.
In Figure \ref{fig1} (left), we show a mixture of 10 spectra with
$E_c$ uniformly distributed from 1 to 3 GeV.  The low energy
(approximately power law) portion of the spectrum is essentially
unchanged from a monoenergetic spectrum with the mean $E_c=2\,$GeV,
while the portion above 2\,GeV is much harder than the single
$E_c=2\,$GeV spectrum.  Figure \ref{fig1} shows that in the case of
this modest mixture the simple PLEC model can underestimate the
spectrum above 10\,GeV by an order of magnitude at the highest
energies accessible to the LAT.

\section{Conclusions and the future}

In summary, the dramatically increased size and breadth of the
$\gamma$-ray pulsar population is finally settling some of the
longstanding questions about the origin and mechanism of $\gamma$-ray
emission.  We now know that $\gamma$ rays arise primarily from the
outer magnetosphere, and that the spectral signature is consistent
with electrons and positrons emitting curvature radiation.  Moreover,
because $\gamma$ rays are such an excellent tracer of the
magnetosphere, detailed analysis of the {\it Fermi} light curves and
spectra offer the opportunity to test more sophisticated emission
theories and probe the magnetosphere structure in model-independent
ways.

Although the trends discussed above are nearly universal, the few
exceptions also offer tantalizing opportunities for better
understanding of the pulsar machine.  The recent detections by VERITAS
and MAGIC (e.g. Aliu et al. 2011, Aleksi{\'c} et al. 2011) of pulsed
photons through energies $>$100\,GeV from the Crab pulsar imply at
least some of the Crab's $\gamma$-ray emission originates from inverse
Compton scattering (see e.g. Lyutikov, Otte, and McCann 2012).
Indeed, of all pulsars detected above 100\,MeV, only the Crab peaks in
$\nu\,F_{\nu}$ well below 1\,MeV, indicating the presence of a strong
synchrotron component and motivating a synchrotron/self-Compton (SSC)
picture for $\gamma$-ray emission.  Discovery and analysis of
``transition'' objects such as PSR B1509-58, whose $\nu\,F_{\nu}$ peak
is of order 1\,MeV (Kuiper et al. 1999, Pilia et al. 2010, den Hartog
et al. in prep), may shed light on the transition, if any, from a
SSC-dominated radiation mechanism to a curvature-dominated one.

\def\nat{{\textit{Nature}}}
\def\apj{{\textit{ApJ}}}
\def\apjs{{\textit{ApJS}}}
\def\mnras{{\textit{MNRAS}}}
\def\aap{{\textit{A\&A}}}

\end{document}